\documentstyle[12pt,epsf]{article}

\newcommand{\eqn}[1]{(\ref{#1})}
\newcommand{\real}{{\bb R}} 
\newcommand{\gstr}{{\gamma_{\rm str}}}  

\font\mybb=msbm10 at 12pt
\def\bb#1{\hbox{\mybb#1}}

\def\e{{\rm e}}
\def\beq{\begin{equation}}
\def\eeq{\end{equation}}
\def\bea{\begin{eqnarray}}
\def\eea{\end{eqnarray}}
\newcommand{\nn}{\nonumber}
\def\bd{\begin{displaymath}}
\def\ed{\end{displaymath}}

\makeatletter
\newdimen\normalarrayskip              
\newdimen\minarrayskip                 
\normalarrayskip\baselineskip
\minarrayskip\jot
\newif\ifold             \oldtrue            
\def\arraymode{\ifold\relax\else\displaystyle\fi} 
\def\@arrayskip{\ifold\baselineskip\z@\lineskip\z@
     \else
     \baselineskip\minarrayskip\lineskip2\minarrayskip\fi}
\def\@arrayclassz{\ifcase \@lastchclass \@acolampacol \or
\@ampacol \or \or \or \@addamp \or
   \@acolampacol \or \@firstampfalse \@acol \fi
\edef\@preamble{\@preamble
  \ifcase \@chnum
     \hfil$\relax\arraymode\@sharp$\hfil
     \or $\relax\arraymode\@sharp$\hfil
     \or \hfil$\relax\arraymode\@sharp$\fi}}
\def\@array[#1]#2{\setbox\@arstrutbox=\hbox{\vrule
     height\arraystretch \ht\strutbox
     depth\arraystretch \dp\strutbox
     width\z@}\@mkpream{#2}\edef\@preamble{\halign \noexpand\@halignto
\bgroup \tabskip\z@ \@arstrut \@preamble \tabskip\z@ \cr}%
\let\@startpbox\@@startpbox \let\@endpbox\@@endpbox
  \if #1t\vtop \else \if#1b\vbox \else \vcenter \fi\fi
  \bgroup \let\par\relax
  \let\@sharp##\let\protect\relax
  \@arrayskip\@preamble}
\makeatother

\newcommand{\newsection}[1]
{\vspace{5mm}
\pagebreak[3]
\addtocounter{section}{1}
\setcounter{equation}{0}
\setcounter{subsection}{0}
\setcounter{footnote}{0}
\begin{flushleft}
{\large\bf \thesection. #1}
\end{flushleft}
\nopagebreak
\medskip
\nopagebreak}

\newlength{\extraspace}
\newlength{\extraspaces}
\begin{document}

\renewcommand{\footnotesize}{\small}

\addtolength{\baselineskip}{.8mm}

\thispagestyle{empty}

\begin{flushright}
\baselineskip=12pt
NBI-HE-99-20\\
OUTP-99-58P\\
hep-th/9910195\\
\hfill{  }\\ Revised April 2000
\end{flushright}
\vspace{.3cm}

\begin{center}

\baselineskip=24pt

{\Large\bf Bottleneck Surfaces and Worldsheet Geometry of Higher-Curvature
Quantum Gravity}\\[15mm]

\baselineskip=12pt

{\bf Richard J. Szabo}\\[3mm]
{\it The Niels Bohr Institute\\ Blegdamsvej 17, DK-2100\\ Copenhagen \O,
Denmark}\\ {\tt szabo@nbi.dk}\\[6mm]{\bf John F. Wheater}\\[3mm]{\it
Department
of Physics -- Theoretical Physics\\ University of Oxford\\ 1
Keble Road, Oxford OX1 3NP, U.K.}\\ {\tt
j.wheater1@physics.oxford.ac.uk}\\[30mm]

{\sc Abstract}

\begin{center}
\begin{minipage}{14cm}

\baselineskip=12pt

We describe a simple lattice model of higher-curvature quantum gravity in two
dimensions and study the phase structure of the theory as a function of the
curvature coupling. It is shown that the ensemble of flat graphs is
entropically unstable to the formation of baby universes. In these simplified
models the growth in graphs exhibits a branched polymer behaviour in the
phase directly before the flattening transition.

\end{minipage}
\end{center}

\end{center}

\vfill
\newpage
\pagestyle{plain}
\setcounter{page}{1}
\stepcounter{subsection}

\newsection{Introduction}

A longstanding problem of quantum gravity is to determine the effects of the
addition of higher-curvature counterterms to the canonical gravity action.
This
problem is most tractable in two-dimensions, mainly because the theory can be
regularized and studied using the discretization approach (see \cite{2drev}
for
reviews). In this case, the simplest term of this type that one could add is
$\kappa\,R(g)^2$, where $R(g)$ is the scalar curvature of a two-dimensional
metric $g$ and the coupling constant $\kappa$ is inversely proportional to the
square of the worldsheet ultraviolet cutoff. The continuum Euclidean partition
function is given by the path integral over all worldsheet metrics
\beq
{\cal Z}=\int Dg~\exp\left[-\int_\Sigma d^2z~\sqrt{\det
g}\,\left(\mu+\frac1{4\pi G}\,R(g)+\kappa\,R(g)^2\right)\right]
\label{partfn}\eeq
where $\mu$ is the cosmological constant and $G$ is the gravitational
constant.
For $\kappa\to0$ the $R^2$ term in the action is irrelevant and the system
lies
in the pure gravitational phase, i.e. that with action consisting of the
cosmological and (topological) Einstein terms and with critical string
exponent
$\gstr=-\frac12$ (for a fixed spherical topology). On the other hand, the
limit
$\kappa\to\infty$ suppresses large curvature fluctuations of the metric and
the
statistical ensemble becomes flatter and flatter at short distance scales. At
$\kappa=\infty$ only flat surfaces contribute to the partition function.

The problem of whether or not this system undergoes a phase transition between
the flat and pure gravitational phases at some finite coupling $\kappa_c$ has
been studied using quantum Liouville theory in \cite{r2calc} and investigated
numerically in \cite{bct}. An exact non-perturbative solution has been
obtained
in \cite{kazr2} for spherical topologies using matrix model techniques and it
was shown that $\gstr=-\frac12$ for all finite values of $\kappa$. The model
always reduces at large length scales to a model of pure gravity, i.e. there
is
no transition to a non-perturbative phase of flat metrics, and only when all
surfaces with non-minimal $R(g)^2$ are completely forbidden ($\kappa=\infty$)
does $\gstr$ change. The key feature of this proof is an appropriate extended
model that interpolates between the fixed lattice (or crystalline) flat space
model and the pure gravity model \cite{fw}. A simplified model which
exhibits the same characteristic features has been studied in \cite{cmw}. In
this paper we shall give a simple demonstration of the absence of a phase
transition from a random to a flat phase of $R^2$ gravity directly in terms of
the fractal structure of two-dimensional quantum gravity. This has the
advantage of being technically much simpler than the solution given in
\cite{kazr2} while at the same time exposing some of the physical and
geometrical characteristics of $R^2$ gravity.

In section 2 we shall describe how to incorporate the effects of the
$R^2$ action in two-dimensions, using the fact that the value of $\gstr$ is
controlled by the structure of baby universe formation on the surfaces of the
ensemble. This leads to the study of surfaces with branching outgrowths
(which we call ``bottleneck surfaces'') which have a contribution to the
 $R^2$ term proportional to the
neck thickness. In section 3 we prove that the generating function for the
bottleneck surfaces is an analytic function of the curvature coupling
$\kappa$,
thereby demonstrating the absence of a phase transition as $\kappa$ is
continuously varied. Finally, in section 4 we present a model which captures
the essential qualitative features of the flattening transition. There we
prove
that the ensemble of flat graphs is unstable to the formation of baby universe
outgrowths, so that the flattening transition can only take place exactly at
$\kappa=\infty$. These results show explicitly that no matter how flat the
system is at short distance scales, it always destabilizes at long wavelengths
into the familiar ensemble of highly fractal baby universes.

\newsection{Baby Universes and Lattice $R^2$ Coupling}

The dynamically triangulated version of pure two-dimensional quantum gravity
is
given by the partition function
\beq
\widehat{Z}(\mu)=\sum_A\e^{-\mu A}\,Z(A)=\sum_A\e^{-\mu A}\,\sum_{{\rm
T}\in{\cal T}_A}1
\label{parttriang}\eeq
where $Z(A)$ is the fixed area partition function and ${\cal T}_A$ is the
topological class of triangulations of area $A$ ($A$ is proportional to the
number of triangles of ${\cal T}_A$). A given choice of ${\rm T}\in{\cal T}_A$
corresponds to a particular discretization of the surface. The local intrinsic
curvature of a vertex $i\in\rm T$ of coordination number $q_i$ is
$R_i=2\pi(6-q_i)/q_i$, so that the discretized form of the $R^2$ action is
given by
\beq
\int_\Sigma d^2z~\sqrt{\det g}~R(g)^2~\longrightarrow~4\pi^2\sum_{i\in\rm
T}\frac{(q_i-6)^2}{q_i}
\label{discrr2}\eeq

However, there is a much simpler way to incorporate the effects of the $R^2$
term by simply finding an appropriate extended model which interpolates
between
pure two-dimensional quantum gravity and a flat phase of worldsheet metrics.
The main idea comes from the fact that the universal constant $\gstr$, which
describes the nature of the geometry, is related to a surface roughness
structure of two-dimensional quantum gravity, the distribution of ``minimal
bottleneck baby universes'' on the ensemble of triangulations \cite{jain}. A
baby universe of area $A_0$ is a small region of the triangulation joined to
the bulk solely by a minimal neck, i.e. a loop consisting of only three links.
Intuitively, they can be pictured as forming  bubblings on the surface. For a
triangulation of area $A\to\infty$ (the continuum limit), the asymptotic
behaviour of the fixed area partition function is
\beq
Z(A)\simeq A^{\gstr-3}~\e^{\mu_0A}
\label{partAasympt}\eeq
where $\mu_0$ is the critical cosmological constant (so that the series
\eqn{parttriang} converges for all $\mu>\mu_0$). From this it can be shown
\cite{jain} that the distribution $n_A(A_0)$ of baby universes is related to
the string susceptibility exponent by
\beq
n_A(A_0)\simeq kA^{3-\gstr}(A-A_0)^{\gstr-2}A_0^{\gstr-2}
\label{babyasympt}\eeq
where here and in the following $k$ denotes a constant which is independent of
the areas and is of the order of unity. This result shows that, as $\gstr$
increases, the number of baby universes increases as $n_A(A_0)/A\simeq
kA_0^{\gstr-2}$. Thus $\gstr$ not only measures the fractal structure of the
typical surfaces that contribute to the partition function, but also their
branching ratio into minimal bottleneck baby universes.

Now consider the situation where the curvature coupling $\kappa$ is very large
and the $\int R^2$ term is thereby trying to make the surfaces as smooth as
possible. A surface of spherical topology has a minimum $\int R^2$, denoted
$R_{\rm min}^2$, which is independent of its area (basically corresponding
to a dual graph consisting of hexagons and 12 pentagons). If we take two such
surfaces (not necessarily of the same area) and join them as shown in fig.~1
by cutting out a triangle from each one and stitching them together, then we
create a surface with one minimum length bottleneck baby universe
and $\int R^2=2R_{\rm min}^2+\Delta$, where $\Delta$ is the contribution to the
$\int R^2$ term arising from the bottleneck. We can iterate this procedure
to produce surfaces with $n$ bottlenecks and
\beq
\int R^2=R_{\rm min}^2+n\left(R_{\rm min}^2+\Delta\right)
\label{latticeintR2}\eeq
The important point here is that $\int R^2$ is not proportional to the area,
but only to the number of bottlenecks. On the other hand, most of the surfaces
of a given area created in this way are geometrically very different from
the single surface with minimum $\int R^2$. We can extend this exercise by
joining surfaces together with larger, non-minimal necks of length $\ell$. 
In this case it is clear that the contribution to $\int R^2$ can be at most
proportional to the length $\ell$ of the bottleneck.

\begin{figure}[htb]
\epsfxsize=4in
\bigskip
\centerline{\epsffile{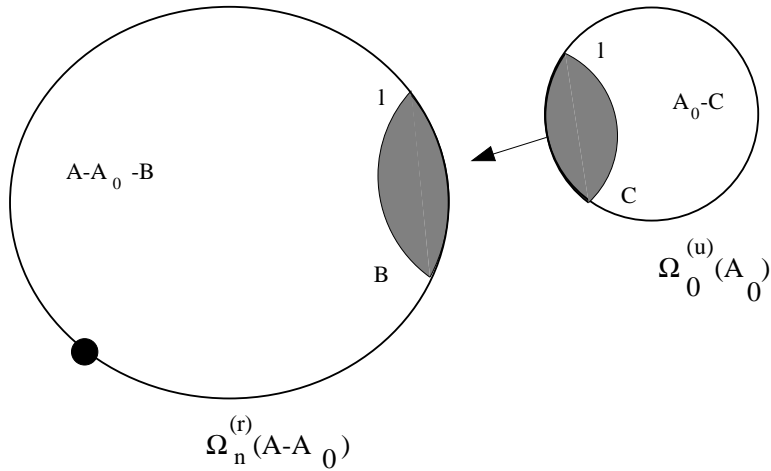}}
\caption{\baselineskip=12pt {\it Gluing an unmarked baby graph of area $A_0$
(counted by $\Omega_0^{(u)}(A_0)$) onto a rooted $n$-bottleneck graph of area
$A-A_0$ (counted by $\Omega_n^{(r)}(A-A_0)$) along a common deleted disk
(shaded areas) to produce a rooted $(n+1)$-bottleneck graph of area $A-B-C$.}}
\bigskip
\label{bottleneck}\end{figure}

To get an idea of how the inclusion of the $R^2$ term modifies
the distribution functions, it is instructive to consider a simple example.
Consider the ensemble of closed rooted trees on a fixed lattice. The total
number of such trees of length $\ell$ (taken to be the number of links in the
tree) is ${\cal N}^{(r)}(\ell)$ which behaves asymptotically as (see
for example \cite{dk})
\beq
{\cal N}^{(r)}(\ell)\simeq\ell^{\alpha_0}~\e^{\rho\ell}
\label{roottrees}\eeq
where $\alpha_0$ is a universal constant which is independent of
the particular type of
 lattice, 
while $\rho>0$ is a non-universal constant. For a given tree of length $\ell$
there are generically $\ell$ different ways of drawing the tree on the
lattice.
We introduce an extra fugacity factor $\e^{-\kappa\ell}$ for each loop, which
is the analog of the curvature term $\int R^2$. The total
number of closed rooted trees on a fixed lattice of area $A$ is then
\beq
{\cal W}(A,\kappa)=\sum_\ell\ell\,{\cal N}^{(r)}(\ell)~\e^{-\kappa\ell}
\label{totaltrees}\eeq
For large areas $A\to\infty$, we can replace the sum in \eqn{totaltrees} by an
integral over $\ell\in[0,\infty)$ and suppress the trees which grow too large
($\ell^2\geq A$) by an exponential damping factor $\e^{-\ell^2/A}$. Thus
asymptotically we have
\beq
{\cal W}(A,\kappa)\simeq\int_0^\infty d\ell~\ell\,{\cal
N}^{(r)}(\ell)~\e^{-\kappa\ell}~\e^{-\ell^2/A}\simeq
kA^{\alpha_0+3/2}~\e^{(\rho-\kappa)^2A/4}
\label{treecont}\eeq
We see that the only effect of the fugacity coupling is to renormalize the
entropy parameter as $\rho\to\rho-\kappa$. The distribution function
\eqn{treecont} is analytic in $\kappa$ and the inclusion of a fugacity factor
for the loops is irrelevant from this point of view. In particular, we may
adjust the coupling so that $\kappa=\rho$, in which case ${\cal W}(A,\rho)$
has
a power law growth with the area. Thus the non-universal tree growth has no
effect on the critical behaviour of the corresponding grand canonical ensemble
quantities either. Note that for a self-avoiding random walk on a regular
square lattice, we have $\alpha_0=-\frac32$ (see for example \cite{dk}),
and ${\cal W}(A,\kappa)$
contains a purely exponential growth with the area $A$ of the lattice. By
tuning to the point $\kappa=\rho$, the distribution in this case is
independent of the size of the lattice.

\newsection{Analytic Structure of the Bottleneck Graph Generating Function}

We will now write down an expression for the partition function of bottleneck
surfaces and argue that it is an analytic function of the $R^2$ coupling
$\kappa$. We work at fixed spherical topology and consider the generating
function for the ensemble of rooted bottleneck graphs which is given by
\beq
\widehat{\Omega}^{(r)}\Bigl(\mu,\kappa;z,\{\alpha_m\}\Bigr)=\sum_A\e^{-\mu
A}\,\sum_nz^n\,\Omega^{(r)}_n\Bigl(A,\kappa;\{\alpha_m\}\Bigr)=\sum_A\e^{-\mu
A}\,\sum_nz^n\,\sum_{{\rm B}\in{\cal B}_A^n}w_{\rm B}
\label{discrgenfn}\eeq
where ${\cal B}_A^n$ denotes the set of all $n$-bottleneck surfaces of area
$A$. The local weight of a graph ${\rm B}\in{\cal B}_A^n$ is given by
\beq
w_{\rm B}=\prod_{b\in{\rm B}}\alpha_{q_b-1}~\e^{-\kappa\ell_b}
\label{graphweight}\eeq
where $b$ are the bottleneck nodes of B of branching number $q_b$ and neck
thickness $\ell_b$, and $\alpha_m$ are arbitrary weights associated with the
nodes. The number $\Omega_n^{(r)}(A)$ of rooted spherical graphs of area $A$
with $n$ bottlenecks (where the area of each triangle is taken to be unity)
can
be constructed inductively as follows. For $n=0$ this number is related to the
number $\Omega_0^{(u)}(A)\equiv Z(A)$ of unmarked spherical graphs as
$\Omega_0^{(r)}(A)=A\,\Omega_0^{(u)}(A)$, since there are $A$ possible
triangles to mark in an unmarked graph. To construct a rooted graph with $n+1$
bottlenecks, we cut out a disk of area $B$ and perimeter $\ell$ from a rooted
graph with $n$ bottlenecks, and a disk of area $C$ with the same boundary
length $\ell$ from a 0-neck unmarked graph (fig. 1). Gluing the two cut graphs
together along the common perimeter of the deleted disks yields a rooted graph
with $n+1$ bottlenecks. In doing so, one must be careful of the degeneracies
which occur in this inductive cutting procedure. Consider two spherical
(0-neck) graphs which are identical to each other {\it except} for their
triangulations inside two disks, of areas $B_0$ and $C_0$ and with identical
boundary length $\ell$, on each respective graph. These two inequivalent
graphs, when glued onto other lattices to produce bottleneck graphs, yield the
{\it same} set of surfaces.

We shall first consider the sub-ensemble of {\it linear} chain bottleneck
graphs. We introduce the disk amplitude $Z_1(\ell;B)$ which is defined to be
the number of surfaces of area $B$ and one boundary of length $\ell$. If we
join two surfaces of areas $B_0$ and $B-B_0>B_0$, each having a single
boundary
of length $\ell$, along their common boundaries, then we obtain a closed
surface of area $B$ with a marked loop of length $\ell$ partitioning it into
two parts of areas $B_0$ and $B-B_0$. Since the boundary has $\ell$ links,
there are generically $\ell$ different ways of joining the two surfaces to
obtain distinct final surfaces (provided that $B_0$ and $B-B_0$ are large
enough). Moreover, any surface of area $B$ with a marked loop of length $\ell$
partitioning it into two parts of areas $B_0$ and $B-B_0$ can be uniquely
constructed in this way. Therefore, for large $B$ and $B_0$, there are
\beq
G_1(B,B_0;\ell)\simeq\ell\,Z_1(\ell;B_0)Z_1(\ell;B-B_0)
\label{Gdef}\eeq
closed surfaces of area $B$ with a marked loop of length $\ell$ that
partitions
the surface into two parts of areas $B_0$ and $B-B_0$. The asymptotic large
area behaviour of the disk amplitude is \cite{mss}
\beq
Z_1(\ell;B)\simeq B^{\gstr-2}~\e^{\mu_0B}\,\ell^{-\gstr-1}~\e^{\rho\ell}
\label{diskasympt}\eeq
where $\rho>0$ is as in \eqn{roottrees}. This amplitude therefore increases
with $\ell$ for small $\ell$, and for large perimeter loops ($\ell^2\geq B$),
$Z_1(\ell;B)$ is exponentially damped by worldsheet finite size
cutoff factors like $\e^{-\ell^2/B}$ \cite{mss}.

We shall also use the annulus amplitude $Z_2(\ell_1,\ell_2;B)$ which is the
number of surfaces of area $B$ with two holes of boundary lengths $\ell_1$ and
$\ell_2$. Its large area asymptotic behaviour is
\beq
Z_2(\ell_1,\ell_2;B)\simeq
B^{\gstr-1}~\e^{\mu_0B}\,(\ell_1+\ell_2)^{-\gstr-1}~\e^{\rho(\ell_1+\ell_2)}
\label{annasympt}\eeq
and for large loops it is suppressed by the worldsheet infrared cutoff
$\e^{-(\ell_1+\ell_2)^2/B}$ \cite{mss}. Then there are
\beq
G_2(B,B_1,B_2;\ell_1,\ell_2)=\ell_1\ell_2\,Z_1(\ell_1;B_1)Z_1(\ell_2;B_2)
Z_2(\ell_1,\ell_2;B-B_1-B_2)
\label{Gadef}\eeq
distinct surfaces of area $B$ with two non-intersecting loops of lengths
$\ell_1$ and $\ell_2$ enclosing areas $B_1$ and $B_2$, respectively.
Neglecting
the graphs counted by $Z_1(\ell;B)$ and $Z_2(\ell_1,\ell_2;B)$ which
themselves
have bottlenecks, i.e. assuming that (\ref{diskasympt}) and (\ref{annasympt})
correctly count the disk and annulus graphs with no bottlenecks of lengths
smaller than $\ell$ and $\ell_1,\ell_2$, an $n$-neck linear chain can then be
constructed by gluing together $n-1$ annulus graphs in between two disk
graphs.
We assume that there is a one-to-one correspondence between the latter
configurations of graphs and those obtained by slicing an $n$-bottleneck graph
along each of its necks. These assumptions will be sufficient to determine the
analytic properties of the full generating function.

We may now write down an expression for the number ${\cal
L}_n^{(r)}(A;\ell_1,\dots,\ell_n)$ of rooted linear graphs of total area $A$
and with $n$ bottlenecks of lengths $\ell_1,\dots,\ell_n$. In the continuum
limit, we can replace sums over areas and bottleneck loop lengths by
integrals.
The lower limits of integration for the perimeter integrals are 3 links (the
minimal bottleneck size), while those of the area integrals are 1 triangle. As
all area integrations are ultraviolet finite, there is no danger in continuing
their limits down to $A_i=0$. Incorporating the smooth infrared cutoffs on
the lengths as described above and the loop fugacity factor
$\e^{-2\kappa\ell_i}$ for each neck, this yields the combinatorial identity
\bea
{\cal
L}_n^{(r)}(A,\kappa;\ell_1,\dots,\ell_n)&=&\left(\prod_{i=1}^{n+1}
\int_0^\infty dA_i\right)~\delta\Bigl(A-
\mbox{$\sum_i$}\,A_i\Bigr)\,A_1\ell_1Z_1(\ell_1;A_1)~
\e^{-2\kappa\ell_1}~\e^{-\ell_1^2/A_1}\nn\\& &\times\,
Z_1(\ell_n;A_{n+1})~\e^{-\ell_n^2/A_{n+1}}
\nn\\& &\times\prod_{i=1}^{n-1}\ell_{i+1}Z_2(\ell_i,\ell_{i+1};
A_{i+1})~\e^{-2\kappa\ell_{i+1}}~\e^{-(\ell_i+\ell_{i+1})^2/A_i}
\label{linearcont}\eea
It is convenient to analyse this relation in the grand canonical ensemble. For
this, we define the Laplace transforms of the loop amplitudes by
\bea
\widehat{Z}_1(\ell;\mu)&\equiv&\int_0^\infty dB~\e^{-\mu
B}\,Z_1(\ell;B)~\e^{-\ell^2/B}\nn\\&\simeq&\frac{\left(\sqrt{\mu-\mu_0}
\,\right)^{1-\gstr}}{\ell^2}~\e^{\rho\ell}\,
K_{1-\gstr}\left(2\sqrt{\mu-\mu_0}\,\ell\right)\nn\\
\widehat{Z}_2(\ell_1,\ell_2;\mu)&\equiv&
\int_0^\infty dB~\e^{-\mu B}\,Z_2(\ell_1,\ell_2;B)~\e^{-(\ell_1+\ell_2)^2/B}
\nn\\&\simeq&\frac{\left(\sqrt{\mu-\mu_0}
\,\right)^{-\gstr}}{\ell_1+\ell_2}~\e^{\rho(\ell_1+\ell_2)}
\,K_{-\gstr}\left(2\sqrt{\mu-\mu_0}\,(\ell_1+\ell_2)\right)
\label{Laplaceloops}\eea
where we have used \eqn{diskasympt} and \eqn{annasympt}, and $K_\nu$ is the
irregular modified Bessel function of order $\nu$. The corresponding
expression
in the grand canonical ensemble is thus
\bea
\widehat{\cal L}_n^{(r)}(\mu,\kappa;\ell_1,\dots,\ell_n)&\equiv&\int_0^\infty
dA~\e^{-\mu A}\,{\cal L}_n^{(r)}(A,\kappa;\ell_1,\dots,\ell_n)\nn\\&=&\frac{
\left(\sqrt{\mu-\mu_0}\,\right)^{1-(n+1)\gstr}}{\ell_1\ell_n}\,
K_{-\gstr}\left(2\sqrt{\mu-\mu_0}\,\ell_1\right)K_{1-\gstr}
\left(2\sqrt{\mu-\mu_0}\,\ell_n\right)\nn\\& &\times~
\e^{2(\rho-\kappa)\sum_i\ell_i}\,
\prod_{i=1}^{n-1}\frac{\ell_i}{\ell_i+\ell_{i+1}}\,
K_{-\gstr}\left(2\sqrt{\mu-\mu_0}\,(\ell_i+\ell_{i+1})\right)
\label{lineargrand}\eea

We are now interested in the analytic properties of the total distribution of
linear $n$-bottleneck graphs
\beq
\widehat{\cal L}_n^{(r)}(\mu,\kappa)\equiv\left(\prod_{i=1}^n\int_3^\infty
d\ell_i\right)~\widehat{\cal L}_n^{(r)}(\mu,\kappa;\ell_1,\dots,\ell_n)
\label{linearmurho}\eeq
as a function of $\kappa\in\real^+$. Because of the ultraviolet cutoff on the
length integrations, possible singularities in (\ref{linearmurho}) would come
only from the behaviour of the integral as $\ell_i\to\infty$. For our purposes
we may therefore approximate the modified Bessel functions in
\eqn{lineargrand}
by their asymptotic behaviours $K_\nu(z)\sim z^{-1/2}~\e^{-z}$ for
$|z|\to\infty$ and consider the function
\bea
\widehat{\cal
L}_n^{(r)}(\mu,\kappa)&\simeq&\frac{\left(\sqrt{\mu-\mu_0}\,
\right)^{-\frac12(n-1)-(n+1)\gstr}}{\left(\sqrt2\,
\right)^{n+1}}\nn\\& &\times\left(\prod_{i=1}^n
\int_3^\infty d\ell_i~\e^{2(\rho-\kappa-2\sqrt{\mu-\mu_0}\,)
\ell_i}\right)~\frac1{\ell_1^{3/2}\ell_n^{3/2}}
\prod_{i=1}^{n-1}\frac{\ell_i}{(\ell_i+\ell_{i+1})^{3/2}}
\label{linearasympt}\eea
In order that \eqn{linearasympt} be generically convergent, we must be in the
phase with $\mu\geq\mu_c$, where
\beq
\mu_c=\mu_0+\frac{(\rho-\kappa)^2}4
\label{critpointrho}\eeq
At $\mu=\mu_c$ there is a phase transition, but the critical behaviour is just
that of the usual continuum limit of the discretized surface model. This will
follow from the fact that both \eqn{linearasympt} and \eqn{critpointrho} are
analytic functions of $\kappa$, and the usual critical point $\mu_c=\mu_0$ can
be reached by tuning the $R^2$ coupling to the value $\kappa=\rho$. We now
focus on the analytic properties of $\widehat{\cal L}_n^{(r)}(\mu,\kappa)$
assuming that $\mu\geq\mu_c$. It can be uniformly bounded by using the
inequality $\ell_i+\ell_{i+1}\geq\sqrt{2\ell_i\ell_{i+1}}$ to write down an
upper bound on the integrations on the right-hand side of \eqn{linearasympt}
which can be evaluated in terms of the incomplete gamma-function
$\Gamma(\alpha,x)=\int_x^\infty dt\,t^{\alpha-1}\,\e^{-t}$ to give a bound on
the derivatives of $\widehat{\cal L}_n^{(r)}(\mu,\kappa)$ with respect to
$\kappa$,
\bea
\left|\frac{\partial^k\widehat{\cal
L}_n^{(r)}(\mu,\kappa)}{\partial\kappa^k}\right|
&\leq&\frac{\left(\sqrt{\mu-\mu_0}\,\right)^{-\frac12(n-1)-(n+1)\gstr}}
{2^{(7n-11)/4-k(n+1)}}\left(2\sqrt{\mu-\mu_0}+\kappa-
\rho\right)^{-(n-5)/2+kn}\nn\\& &\times\,\Gamma
\left(-\mbox{$\frac14$}\,+k,6(2\sqrt{\mu-\mu_0}+\kappa-
\rho)\right)\Gamma\left(-\mbox{$\frac54$}\,+k,6
(2\sqrt{\mu-\mu_0}+\kappa-\rho)\right)\nn\\& &\times\,
\Gamma\left(\mbox{$\frac12$}\,+k,6(2\sqrt{\mu-\mu_0}
+\kappa-\rho)\right)^{n-2}
\label{linearbound}\eea
for $k\geq0$.

For non-integer or negative $\alpha$, $\Gamma(\alpha,x)$ is a multi-valued
function of $x$ with a branch cut along the negative real $x$-axis. In the
phase with $\mu>\mu_c$, the linear $n$-bottleneck distribution is therefore
uniformly bounded by an analytic function of $2\sqrt{\mu-\mu_0}+\kappa-\rho$.
In terms of the generating function for the linear bottleneck graph ensemble,
defined as a formal power series
\beq
\widehat{\cal L}^{(r)}(\mu,\kappa;z)=\sum_{n=0}^\infty z^n\,\widehat{\cal
L}_n^{(r)}(\mu,\kappa)
\label{lineargenfn}\eeq
in a variable $z$, this bound reads
\bea
\left|\frac{\partial^k\widehat{\cal
L}^{(r)}(\mu,\kappa;z)}{\partial\kappa^k}\right|&\leq&
\frac{\Gamma\left(-\mbox{$\frac14$}\,+k,6(2\sqrt{\mu-\mu_0}+
\kappa-\rho)\right)\Gamma\left(-\mbox{$\frac54$}\,
+k,6(2\sqrt{\mu-\mu_0}+\kappa-\rho)\right)}{\Gamma
\left(\mbox{$\frac12$}\,+k,6(2\sqrt{\mu-\mu_0}+\kappa-
\rho)\right)^2}\nn\\& &\times\,2^{2k+3}\,\left(\sqrt{\mu-\mu_0}\,
\right)^{\frac12-\gstr}\left(2\sqrt{\mu-\mu_0}+\kappa-\rho
\right)^{5/2}\nn\\& &\times\left[2^{k+1/4}-4z\,
\left(\sqrt{\mu-\mu_0}\,\right)^{-\frac12-\gstr}
\left(2\sqrt{\mu-\mu_0}+\kappa-\rho\right)^{-1/2+k}\right.
\nn\\& &\biggl.\times\,\Gamma\left(\mbox{$\frac12$}\,
+k,6(2\sqrt{\mu-\mu_0}+\kappa-\rho)\right)\biggr]^{-1}
\label{lineargenfnbound}\eea
Being uniformly bounded by an analytic function, the generating function
cannot
have any singularities as $\kappa>\rho-2\sqrt{\mu-\mu_0}$ is varied and it is
analytic in the variable $2\sqrt{\mu-\mu_0}+\kappa-\rho$. Furthermore, since
the critical cosmological constant \eqn{critpointrho} is itself an analytic
function of $\kappa$, there are no singularities as $\kappa$ is varied
throughout its range. Therefore, the linear bottleneck generating function is
an analytic function of $\kappa\in\real^+$. There are no phase transitions
within the ensemble of linear bottleneck graphs as one continuously varies the
$R^2$ coupling constant.

We now consider the full bottleneck graph ensemble with generating function
\eqn{discrgenfn}. To study a bottleneck distribution with arbitrary
branchings,
we shall need to use the genus 0 $n$-loop amplitude
$Z_n(\ell_1,\dots,\ell_n;B)$. We can determine its approximate asymptotic
behaviour as follows. For each additional loop that is drawn on a graph there
is an extra combinatorial area factor giving an overall entropy $B^nZ(B)$ for
locating the centers of the loops on the closed surface. Furthermore, each
loop
on the surface contributes an exponential length growth $\e^{\rho\ell_i}$, and
the smooth infrared cutoff on large perimeter loops is
$\e^{-(\sum_i\ell_i)^2/B}$. The crucial point, however, is that the $n$-loop
amplitude depends only on the sum of the loop lengths \cite{mss}. The natural
ansatz is then
\beq
Z_n(\ell_1,\dots,\ell_n;B)\simeq
B^{\gstr-3+n}~\e^{\mu_0B}\left(\sum_{i=1}^n\ell_i
\right)^{-\gstr-1}~\e^{\rho\sum_i\ell_i}~\e^{-(\sum_i\ell_i)^2/B}
\label{nloopasympt}\eeq
We note that this ansatz only dictates the large area and loop length
dependence of the amplitude, but not its coefficient which depends on $n$.

To establish \eqn{nloopasympt}, we use induction on $n$ and the consistency
condition
\beq
Z_{n-1}(\ell_1,\dots,\ell_{n-1};B)=\int_1^BdB_0~\int_3^\infty
d\ell~Z_n(\ell_1,\dots,\ell_{n-1},\ell;B_0)Z_1(\ell;B-B_0)
\label{nloopconscond}\eeq
which represents the combinatorics of gluing together disk amplitudes and
$n$-loop amplitudes to generate $(n-1)$-loop graphs. Using the ansatz
\eqn{nloopasympt} and changing variables in the area integral to
$x=B_0/(B-B_0)$, the right-hand side of \eqn{nloopconscond} becomes
\bea
& &B^{2\gstr-4+n}~\e^{\mu_0B}~\e^{\rho L_{n-1}}~\e^{-L_{n-1}^2/B}\int_3^\infty
d\ell~\ell^{-\gstr-1}\,(\ell+L_{n-1})^{-\gstr-1}~
\e^{2\rho\ell-2\ell(\ell+L_{n-1})/B}\nn\\& &~~~~~~
\times\int_{-1}^\infty dx~x^{\gstr-3+n}\,(x+1)^{-2\gstr+3-n}~
\e^{-\ell^2x/B-(\ell+L_{n-1})^2/Bx}
\label{varchange}\eea
where $L_{n-1}=\sum_{i=1}^{n-1}\ell_i$. For $B\to\infty$ we may approximate
the
integral over $x$ in \eqn{varchange} by its asymptotic behaviour which
integrates to the modified Bessel function
\beq
2\left(\frac{\ell+L_{n-1}}\ell\right)^{\gstr-1}\,K_{\gstr-1}
\Bigl(2\ell(\ell+L_{n-1})/B\Bigr)
\label{xBessel}\eeq
{}From the asymptotic behaviour
$K_\nu(x)\simeq2^{\nu-1}\Gamma(\nu)\,x^{-\nu}+\dots$ for $x\to0$ and
$\nu\neq0$, it follows that the large-area behaviour of \eqn{varchange} is
thus
\beq
B^{3\gstr-5+n}~\e^{\mu_0B}~\e^{\rho L_{n-1}}~\e^{-L_{n-1}^2/B}\int_3^\infty
d\ell~\ell^{-\gstr+1}\,(\ell+L_{n-1})^{-\gstr-1}~
\e^{2\rho\ell-2\ell(\ell+L_{n-1})/B}
\label{lastloopint}\eeq
Within the present approximations the loop integral \eqn{lastloopint} may be
evaluated using the saddle-point approximation. The stationary condition is
\beq
2\left(\rho-\frac{L_{n-1}}B\right)\ell(\ell+L_{n-1})-
\frac{4\ell^2(\ell+L_{n-1})}B-(\gstr+1)\ell-(\gstr-1)(\ell+L_{n-1})=0
\label{saddleeqn}\eeq
Solving the cubic equation \eqn{saddleeqn} and evaluating the one-loop
fluctuation integral corresponding to \eqn{lastloopint} for the root which is
positive and regular at $L_{n-1}=0$, we find in the limit $B\to\infty$ that
\eqn{lastloopint} is proportional to the left-hand side of \eqn{nloopconscond}
as given by \eqn{nloopasympt}. Taking the Laplace transform of the
expression \eqn{nloopasympt}, we find the corresponding amplitude in the grand
canonical ensemble,
\bea
\widehat{Z}_n(\ell_1,\dots,\ell_n;\mu)&\simeq&\left(\sqrt{\mu-\mu_0}
\,\right)^{2-n-\gstr}\left(\sum_{i=1}^n\ell_i\right)^{n-3}~
\e^{\rho\sum_i\ell_i}\,
K_{2-n-\gstr}\left(2\sqrt{\mu-\mu_0}
{}~\mbox{$\sum_i\ell_i$}\right)\nn\\&\sim&\left(\sqrt{\mu-\mu_0}
\,\right)^{\frac32-n-\gstr}\left(\sum_{i=1}^n\ell_i
\right)^{n-\frac72}~\e^{(\rho-2\sqrt{\mu-\mu_0}\,)\sum_i\ell_i}
{}~~~~~~{\rm for}~~\ell_i\to\infty\nn\\& &
\label{nlooplaplace}\eea

With the bottleneck node couplings $\alpha_m$ completely arbitrary, the full
generating function may be generated by the recursive equation which is
represented symbolically in fig. 2. Summing over all configurations shown
there
gives the identity
\bea
& &\int_0^\infty dA_1~dA_2~\delta(A-A_1-A_2)\int_3^\infty
d\ell~\ell~\e^{-\kappa\ell}\,Z_1^{(r)}(\ell;A_1)\,
\Omega\Bigl(\ell,A_2,\kappa;z,\{\alpha_m\}\Bigr)\nn\\& &=z
\left[A\,Z(A)+\sum_{n=1}^\infty\alpha_n\left(
\prod_{i=1}^{n+1}\int_0^\infty dA_i\right)~\delta
\Bigl(A-\mbox{$\sum_i$}\,A_i\Bigr)\left(\prod_{i=1}^n
\int_3^\infty d\ell_i~\ell_i~\e^{-\kappa\ell_i}\right)\right.
\nn\\& &~~~~~~\left.\times\,Z_n^{(r)}(\ell_1,\dots,\ell_n;A_1)\,
\prod_{i=1}^n\Omega\Bigl(\ell_i,A_{i+1},\kappa;z,\{\alpha_m\}\Bigr)\right]
\label{genfnid}\eea
where $Z_n^{(r)}(\ell_1,\dots,\ell_n;A)=A\,Z_n(\ell_1,\dots,\ell_n;A)$ is the
marked $n$-loop amplitude and we have introduced the generating function
$\Omega(\ell,A,\kappa;z,\{\alpha_m\})$ for bottleneck surfaces of area $A$ and
with a single deleted disk of boundary length $\ell$. The left-hand side of
\eqn{genfnid} is the desired quantity $\Omega^{(r)}(A,\kappa;z,\{\alpha_m\})$
and from \eqn{nlooplaplace} it follows that the corresponding relationship in
the grand canonical ensemble is
\bea
\widehat{\Omega}^{(r)}\Bigl(\mu,\kappa;z,\{\alpha_m\}\Bigr)
&\equiv&\int_3^\infty
d\ell~\ell~\e^{(\rho-\kappa)\ell}\,K_{-\gstr}\left(2\sqrt{\mu-\mu_0}\,
\ell\right)\widehat{\Omega}\Bigl(\ell,\mu,\kappa;z,\{\alpha_m\}\Bigr)\nn\\&=&z
\left[(\mu-\mu_0)^{1-\gstr}+\sum_{n=1}^\infty\alpha_n\left(\sqrt{\mu-\mu_0}\,
\right)^{1-n-\gstr}\right.\nn\\& &\times\prod_{i=1}^n\int_3^\infty
d\ell_i~\ell_i~\e^{(\rho-\kappa)\ell_i}\,\widehat{\Omega}
\Bigl(\ell_i,\mu,\kappa;z,\{\alpha_m\}\Bigr)\nn\\& &\left.\times
\left(\sum_{i=1}^n\ell_i\right)^{n-2}\,K_{1-n-\gstr}
\left(2\sqrt{\mu-\mu_0}\,\mbox{$\sum_i$}\,\ell_i\right)\right]
\label{genfnidhat}\eea
This iterative relation defines a complicated integral equation for
$\widehat{\Omega}(\ell,\mu,\kappa;z,\{\alpha_m\})$ whose solution gives the
generating function \eqn{discrgenfn}. The complexity of the identity
\eqn{genfnidhat} makes an explicit solution or even an analyticity analysis
intractable. However, one may argue that the solution to \eqn{genfnidhat} is
an
analytic function of $\kappa$ as follows. A generic branching distribution
function for the bottleneck ensemble will always involve products of functions
of the form \eqn{nlooplaplace}. From the inequalities
\beq
\left(\sum_{i=1}^n\ell_i\right)^{n-\frac72}\leq\left\{
{\begin{array}{rrl}n^{(n-\frac72)/n}\,
\prod_i\ell_i^{(n-\frac72)/n}~~~~&,&~~~~n<4
\\n^{n-\frac72}\,\max_i\ell_i^{n-\frac72}~~~~&,&~~~~n\geq4\end{array}}\right.
\label{nloopboundid}\eeq
it follows that a general bottleneck function can always be bounded from above
by a function which is given by a product of incomplete gamma-functions and
other elementary functions. In the phase $\mu>\mu_c$, the $n$-neck
distribution
functions are analytic in $\rho$. The grand canonical distribution function
$\widehat{\Omega}^{(r)}(\mu,\kappa;z)$ is therefore expected to be an analytic
function of $\kappa\in\real^+$. Note that the recursive definition
\eqn{genfnidhat} is reminescent of that for the generating function of a
branched polymer ensemble \cite{bp}. In the next section we shall consider a
slight simplification of the model defined by \eqn{genfnidhat} which is
amenable to explicit analysis and thereby demonstrate that this similarity is
not a coincidence.

\begin{figure}[htb]
\epsfxsize=4in
\bigskip
\centerline{\epsffile{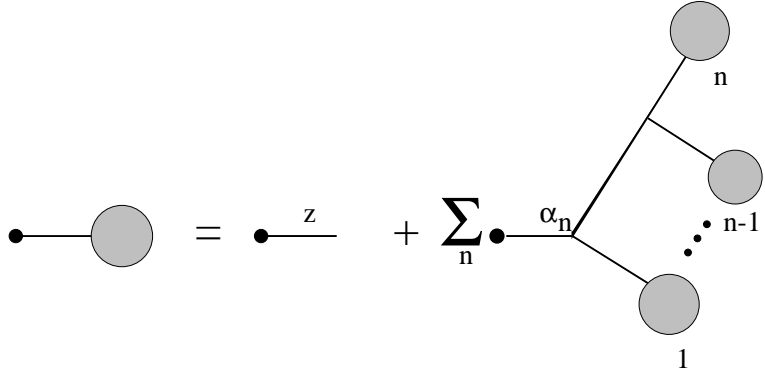}}
\caption{\baselineskip=12pt {\it Schematic representation of the iterative
definition of the bottleneck surface generating function. The shaded circles
represent the loop functions $\Omega(\ell,A,\kappa;z,\{\alpha_m\})$ while the
lines represent $0$-neck surfaces. Each attachment is done by gluing the
surfaces together along a common deleted disk.}}
\bigskip
\label{iterative}\end{figure}

\newsection{A Model for the Flattening Transition}

We shall now present a model for the transition between flat and random graphs
which captures the essential qualitative features of the flattening
transition.
We will consider two simplifications of the full bottleneck graph generating
function. First, we start with the ensemble of regular square
lattices,\footnote{Actually, one must consider square lattices with positive
curvature defect insertions to be able to close them on a spherical topology.
This modification would not affect the qualitative behaviour described in the
following. The model could of course be extended to include non-square
lattices
and arbitrarily shaped bottlenecks.} which in the schematic representation of
fig. 2 are represented by
the rooted lines. Let ${\cal Z}_\ell^{(0)}(A)$ be the number of such flat
graphs of area $A$ and with a single square loop of side $\ell$ drawn on them
(the
0-bottleneck partition function). We are interested in the generating function
${\cal Z}_\ell(A)$ for the ensemble of bottleneck surfaces of area $A$ with a
single loop of side $\ell$ drawn on them, which is depicted by the shaded
circles in fig. 2. It is constructed recursively as described at the end of
the
previous section. We start with a flat base lattice of area $A_0$ and put an
outgrowth
of bottleneck graphs with a square loop of side $n$ on it. There are $A_0$
possible
ways to do this and the area of the new graph is $A_0-n^2$. However, the
effective area remaining to insert a second outgrowth of bottleneck surfaces
of
loop perimeter $4n$ is $A_0-4n^2$ because of the excluded area effect. Each
such
outgrowth is attached to the base graph with a fugacity $\e^{-\kappa w(n)}$,
where $w(n)$ is a function which satisfies $w(1)=1$ and
$w'(n)\ge 0$, but which does not grow as fast as $n^2$.

This procedure can be used to build the full generating function ${\cal
Z}_\ell(A)$ by summing over all possible ways of attaching outgrowths to the
base graph. Dividing by the permutation symmetry factor for a $k$-branch
surface, we have
\bea
{\cal Z}_\ell(A,\kappa)&=&{\cal Z}_\ell^{(0)}(A)+\sum_{A_0=\ell^2}^\infty{\cal
Z}_\ell^{(0)}(A_0)\sum_{n=1}^{\frac12\sqrt{A_0}}\,
\sum_{k=1}^{A_0/4n^2}\frac1{k!}\nn\\& &\times\left[
\prod_{i=1}^k\left(A_0-4(i-1)n^2\right)\sum_{A_i=n^2}^\infty
{\cal Z}_n(A_i,\kappa)~\e^{-\kappa w(n)}\right]\,\delta
\left(A-A_0+kn^2-\mbox{$\sum_i$}\,A_i\right)\nn\\& &
\label{Zlgen}\eea
It is again convenient to analyse the grand canonical partition function
which here is defined by the discrete Laplace transform
\bea
\widehat{\cal Z}_\ell(\mu,\kappa)&\equiv&\sum_{A=\ell^2}^\infty\e^{-\mu
A}\,{\cal Z}_\ell(A,\kappa)\nn\\&=&\widehat{\cal
Z}_\ell^{(0)}(\mu)+\sum_{A=\ell^2}^\infty\sum_{n=1}^{\frac12\sqrt A}\e^{-\mu
A}\,{\cal Z}_\ell^{(0)}(A)\left[\left(1+4n^2~\e^{\mu n^2-\kappa
w(n)}\,\widehat{\cal
Z}_n(\mu,\kappa)\right)^{A/4n^2}-1\right]\nn\\& &
\label{Zlmu}\eea
where we have used the binomial theorem to do the sum over $k$ from
\eqn{Zlgen}.
Upon carefully interchanging the sums over $n$ and $A$ in \eqn{Zlmu} we arrive
at
\beq
\widehat{\cal Z}_\ell(\mu,\kappa)=\widehat{\cal
H}_\ell^{(0)}(\mu)+\sum_{n=1}^{\ell-1}\widehat{\cal
Z}_\ell^{(0)}(\bar\mu_n)+\sum_{n=\ell}^\infty\widehat{\cal
Z}_n^{(0)}(\bar\mu_n)
\label{Zlrecrel}\eeq
where we have introduced the function
\beq
\widehat{\cal H}_\ell^{(0)}(\mu)=\widehat{\cal
Z}_\ell^{(0)}(\mu)-\frac12\sum_{A=\ell^2}^\infty\sqrt A~\e^{-\mu
A}\,{\cal
Z}_\ell^{(0)}(A),
\label{Gldef}\eeq
and the rescaled cosmological constants
\beq
\bar\mu_n(\mu,\kappa)=\mu-\frac1{4n^2}\log\left(1+4n^2~\e^{\mu
n^2-\kappa w(n)}\,\widehat{\cal Z}_n(\mu,\kappa)\right).
\label{barmudef}\eeq
 Note that the grand canonical base graph generating function
behaves asymptotically for large loops as
\beq
\widehat{\cal
Z}_\ell^{(0)}(\mu)=\e^{-\mu\ell^2}\,g_\ell(\mu)\label{Zl0asympt}\eeq
where $g_\ell(\mu)< g_1(\mu)$ and $g_1(\mu)$ converges for all positive
$\mu$ and diverges at $\mu=0$. It follows that the sums in \eqn{Zlrecrel}
converge provided $\bar\mu_n(\mu,\kappa)>0$.

We will now analyse the analytic properties of the solution of the
infinite-term recursion relation \eqn{Zlrecrel} as a function of $\mu$ and
$\kappa$. From \eqn{Zlmu} it follows that
\beq \widehat{\cal Z}_\ell(\mu,\kappa=\infty)=\widehat{\cal Z}_\ell^{(0)}(\mu)
\label{flatlimit}\eeq
so that the suppression of curvature fluctuations via an infinite
$R^2$-coupling leaves only flat (square lattice) graphs. In this flat phase,
the partition function diverges only at the critical point $\mu_c^{(0)}=0$ and
formally the string susceptibility exponent is $\gamma_{\rm str}^{(0)}=2$. For
$\kappa<\infty$, it follows immediately from \eqn{Zlrecrel} that the partition
function $\widehat{\cal Z}_\ell(\mu,\kappa)$ determines a different
universality class than the flat phase. To see this, we suppose that
$\widehat{\cal Z}_\ell(\mu,\kappa)$ has the {\it same} critical behaviour as
$\widehat{\cal Z}_\ell^{(0)}(\mu)$, i.e. that $\widehat{\cal
Z}_\ell(\mu,\kappa)$ diverges at some critical point $\mu=\mu_c$.
If this were the case then the smallest of the $\bar\mu_n(\mu,\kappa)$ in
\eqn{barmudef} would become zero at some $\mu=\mu_c'>\mu_c$, and so from
\eqn{Zlrecrel} and \eqn{Zl0asympt}
 the partition function
$\widehat{\cal Z}_\ell(\mu,\kappa)$ diverges before its critical point is
reached. Thus a critical behaviour at {\it any} $\kappa<\infty$ whereby
$\widehat{\cal Z}_\ell(\mu,\kappa)$ itself diverges is inconsistent. It is
only
at $\kappa=\infty$ that the universality class of the random geometry changes.

Physically then, the flat lattice system at $\kappa=\infty$ is
entropically unstable to the formation of baby universes that make up the full
gravitational ensemble, which gives a worldsheet geometric picture of why
there
is no flattening phase transition in the two-dimensional $R^2$ quantum gravity
model. In the remainder of this paper we will deduce what the nature is of the
phase when one perturbs the flat graph ensemble by baby universe outgrowths as
described above. To get an idea of what the system at $\kappa<\infty$
represents,
consider the simplification whereby we allow outgrowths of only a single loop
side $\ell$. This means that we keep only the $n=\ell$ term in \eqn{Zlgen} and
proceeding as in the general case we would then arrive at
\beq
\widehat{\cal Z}_\ell(\mu,\kappa)=\widehat{\cal Z}_\ell^{(0)}(\bar\mu_\ell).
\label{BPeqn}\eeq
This is the standard behaviour of the generating function for a branched
polymer ensemble \cite{durhuus}. Indeed, as before, $\widehat{\cal
Z}_\ell(\mu,\kappa)$ cannot diverge at the critical point $\mu_c$ because
then $\bar\mu_\ell(\mu,\kappa)$ would reach 0 before $\widehat{\cal
Z}_\ell(\mu,\kappa)$
reaches $\infty$. Differentiating both sides of \eqn{BPeqn} with respect to
$\mu$ using \eqn{barmudef} yields
\beq
\frac{\partial\widehat{\cal
Z}_\ell(\mu,\kappa)}{\partial\mu}=\frac{\left(1-3\ell^2~
\e^{\mu\ell^2-\kappa w(\ell)}\,\widehat{\cal Z}_\ell(\mu,\kappa)\right)
\e^{-4\ell^2(\mu-\bar\mu_\ell)}\,\frac{\partial\widehat{\cal
Z}_\ell^{(0)}(\bar\mu_\ell)}{\partial\bar\mu_\ell}}
{1+\e^{-\ell^2(3\mu-4\bar\mu_\ell)
-\kappa w(\ell)}\,\frac{\partial\widehat{\cal Z}_\ell^{(0)}(\bar\mu_\ell)}
{\partial\bar\mu_\ell}}.
\label{BPeqnderiv}\eeq
Since $\partial\widehat{\cal
Z}_\ell^{(0)}(\bar\mu_\ell)/\partial\bar\mu_\ell\to-\infty$ as
$\bar\mu_\ell\to0$, it follows that the denominator of \eqn{BPeqnderiv}
vanishes at some finite value of $\bar\mu_\ell$ where the numerator is regular
and non-vanishing. Thus $\partial\widehat{\cal
Z}_\ell(\mu,\kappa)/\partial\mu$
diverges at the critical point, and generically we get a branched polymer
ensemble with string susceptibility exponent $\gstr=+\frac12$ \cite{durhuus}.
However, the actual bottleneck ensemble contains a complicated mixing of all
loop lengths and more care must be exercised
 in deducing the critical behaviour.

 We begin by deducing some analytic
properties of $\widehat{\cal Z}_\ell(\mu,\kappa)$ for
$\kappa<\infty$ using the recursion relation \eqn{Zlrecrel}. Note
that for fixed $\ell$, $\widehat{\cal
Z}_\ell(\mu,\kappa)$ is a monotonic decreasing function of $\mu\in\real^+$,
while for fixed $\mu$ it is a decreasing function of $\ell$. Let us
examine the behaviour of the functions $\bar\mu_n$ very close to the
critical point $\mu=\mu_c$. Since $\mu_c=0$ at $\kappa=\infty$, at very large
$\kappa$ we expect that $\mu_c$ is very small so that for
$n^2\mu\ll1$ we  find
\bea
\bar\mu_{n+1}-\bar\mu_n&>&\frac1{4n^2}\log\left(1+4n^2~\e^{\mu
n^2-\kappa w(n)}\,\widehat{\cal Z}_n(\mu,\kappa)\right)\nn\\&
&-\frac1{4(n+1)^2}\log\left(1+4(n+1)^2~\e^{\mu(n+1)^2-\kappa
w(n+1)}\,\widehat{\cal
Z}_n(\mu,\kappa)\right)\nn\\&\simeq&\frac1{4n^2}\log\left(1+4n^2~
\e^{-\kappa w(n)}\,\widehat{\cal Z}_n(\mu_c,\kappa)\right)
\nn\\& &-\frac1{4(n+1)^2}\log\left(1+4(n+1)^2~\e^{-\kappa w(n+1)}\,
\widehat{\cal Z}_n(\mu_c,\kappa)\right)\nn\\&>&0
\label{barmuincr}\eea
for $\mu\sim\mu_c$ and $\kappa\gg1$. Thus 
$\{\bar\mu_n\}$ is an  increasing sequence for small $n$. Denoting
$\bar\mu_{\rm min}=\inf_n\bar\mu_n$, it follows from \eqn{Zlmu} and
\eqn{Zl0asympt} that
\bea
\widehat{\cal Z}_\ell(\mu,\kappa)&\leq&\e^{-\bar\mu_{\rm min}\ell^2}\,
g_1(\bar\mu_{\rm min})\left(\ell+\frac{1}{2}\,\sqrt{\frac{\pi}{\bar\mu_{\rm
min}}}\,\right).
\label{Zln0lowerbd}\eea
 On the other hand, because $\bar\mu_{\rm min}<\mu$, from \eqn{Zlrecrel} it
follows that for large 
$\ell$ we have
\beq
\widehat{\cal Z}_\ell(\mu,\kappa)  > \e^{-\bar\mu_{\rm min}\ell^2}
\label{Zln0upperbd}\eeq
The bounds \eqn{Zln0lowerbd} and \eqn{Zln0upperbd} show that for large $\ell$
we may write
\beq
\widehat{\cal Z}_\ell(\mu,\kappa)\simeq\e^{-\bar\mu_{\rm
min}\ell^2}\,f(\ell,\mu)
\label{Zlapprox}\eeq
where the function $f(\ell,\mu)$ is larger than 1  and
 grows at most linearly with $\ell$.  We then see 
that at large enough $n$ 
\beq \bar\mu_{n+1}-\bar\mu_n\simeq \frac{\log 4n^2}{4n^2}
-\frac{\log 4(n+1)^2}{4(n+1)^2}
+\frac{\kappa w(n+1)}{4(n+1)^2}-\frac{\kappa w(n)}{4n^2}\eeq
so that now, depending upon the nature of the function $w(n)$, the
$\bar\mu_n$ may start to fall again. However, we also 
have that 
\beq
\bar\mu_\infty\equiv\lim_{n\to\infty}\bar\mu_n=\frac34\mu+\frac14\bar\mu_{\rm
min}
>\bar\mu_{\rm min}
\label{muinfty}\eeq
so that we can identify $\bar\mu_{\rm min}=\bar\mu_1$.
It follows from these properties of the sequence $\{\bar\mu_n\}$ that the
infinite series in
\eqn{Zlrecrel} is absolutely convergent.

We now rewrite \eqn{Zlrecrel} in the form
\beq
\widehat{\cal Z}_\ell(\mu,\kappa)=\widehat{\cal Z}_1(\mu,\kappa)+\widehat{\cal
H}_\ell^{(0)}(\mu)-\widehat{\cal
H}_1^{(0)}(\mu)+\sum_{n=1}^{\ell-1}\left(\widehat{\cal
Z}_\ell^{(0)}(\bar\mu_n)-
\widehat{\cal Z}_n^{(0)}(\bar\mu_n)\right)
\label{recrelrewrite}\eeq
The right-hand side of \eqn{recrelrewrite} depends only on the known functions
 $\widehat{\cal
H}_\ell^{(0)}(\mu)$ and the partition functions $\widehat{\cal
Z}_n(\mu,\kappa)$ for
$1\leq n<\ell$. Iterating \eqn{recrelrewrite} thus
determines $\widehat{\cal Z}_n(\mu,\kappa)$ as a function only of $\mu$,
$\kappa$, and
$\widehat{\cal Z}_1$, for any $n>1$. Using \eqn{Zlrecrel} we see that
 $\widehat{\cal Z}_1$ satisfies
\beq
\widehat{\cal Z}_1(\mu,\kappa)=\widehat{\cal H}_1^{(0)}(\mu)+{\cal
F}\left(\mu,\kappa,\widehat{\cal Z}_1(\mu,\kappa)\right)
\label{Z1eqn}\eeq
where implicitly the function $\cal F$ is given by
\bea
{\cal F}\left(\mu,\kappa,\widehat{\cal
Z}_1\right)&=&\sum_{n=1}^\infty\widehat{\cal Z}_n^{(0)}(\bar\mu_n)\\
\bar\mu_n(\mu,\kappa,\widehat{\cal
Z}_n)&=&\mu-\frac1{4n^2}\log\left(1+4n^2~\e^{\mu
n^2-\kappa w(n)}\,\widehat{\cal Z}_n\right)
\label{calFdef}\eea
together with  \eqn{recrelrewrite},
but explicitly it is a very complicated function. Moreover, \eqn{calFdef} is
an
analytic function of $\mu$, $\kappa$, and $\widehat{\cal Z}_1$. 

 Differentiating \eqn{Z1eqn} with respect to $\mu$ gives
\beq
\left.\frac{\partial\widehat{\cal
Z}_1}{\partial\mu}\right)_\kappa=\frac{\frac{d\widehat{\cal
H}_1^{(0)}}{d\mu}+\left.\frac{\partial{\cal
F}}{\partial\mu}\right)_{\kappa,\widehat{\cal
Z}_1}}{1-\left.\frac{\partial{\cal F}}{\partial\widehat{\cal
Z}_1}\right)_{\mu,\kappa}}.
\label{dZ1eqn}\eeq
The numerator of \eqn{dZ1eqn} is a well behaved function for $\mu>0$ but at 
some critical point the denominator will vanish and thereby generate a
non-analyticity
in $\widehat{\cal Z}_1$. To see this, we note from \eqn{calFdef} that
\beq \left.\frac{\partial{\cal F}}{\partial\widehat{\cal
Z}_1}\right)_{\mu,\kappa}=\sum_{n=1}^\infty{\cal A}^{n,n}
f_{n}\label{BPcondn}
\eeq
where we define 
\beq{\cal
A}^{\ell,n}=-\widehat{\cal Z}_\ell^{(0)\prime}(\bar\mu_n)
~\e^{n^2(4\bar\mu_n-3\mu)-\kappa w(n)}\label{calAdef}\eeq
and
\beq f_{\ell}\equiv\left.\frac{\partial\widehat{\cal
Z}_\ell}{\partial\widehat{\cal Z}_1}\right)_{\mu,\kappa}=1+
\sum_{n=1}^{\ell-1}\left({\cal A}^{\ell,n}-{\cal A}^{n,n}\right)\,f_{n}
\label{flrecrel}\eeq
where we have used \eqn{recrelrewrite}.
Note that ${\cal A}^{\ell,n}$ is a monotonically decreasing function of
$\ell$,
so that \eqn{flrecrel} implies
\beq
f_{\ell+1}-f_\ell=\sum_{n=1}^\ell\left({\cal A}^{\ell+1,n}-{\cal
A}^{\ell,n}\right)\,f_n<0
\label{fldecr}\eeq
provided that $f_n>0~~\forall n\geq1$.
 Thus $\{f_\ell\}$ is a monotonically decreasing sequence. Under
these same assumptions we also have
\bea
\sum_{n=1}^{\ell-1}{\cal
A}^{\ell,n}\,f_n&=&\sum_{n=1}^{\ell-1}f_n\Bigl(\ell^2g_\ell(\bar\mu_n)
-g_\ell^{\prime}(\bar\mu_n)
\Bigr)~\e^{-3n^2(\mu-\bar\mu_n)-\kappa w(n)}~\e^{(n^2-\ell^2)\bar\mu_n}
\nn\\&<&\e^{-(2\ell-1)\bar\mu_1}\sum_{n=1}^{\ell-1}
f_n\Bigl(\ell^2g_\ell(\bar\mu_n)-g_\ell^{\prime}(\bar\mu_n)
\Bigr)~\e^{-3n^2(\mu-\bar\mu_n)-\kappa w(n)}.
\label{sumAfnbd}\eea
The last line of \eqn{sumAfnbd} vanishes in the limit $\ell\to\infty$, so that
the first series in the recursion relation \eqn{flrecrel} vanishes as
$\ell\to\infty$. Using this fact and \eqn{flrecrel} it follows that
\eqn{BPcondn} can be written in terms of the single quantity
$f_\infty=\lim_{\ell\to\infty} f_\ell$ as
\beq
\frac{\partial{\cal F}}{\partial\widehat{\cal Z}_1}=1-f_\infty.
\label{dFZ1f}\eeq
Again assuming that the sequence $\{f_\ell\}$ is  monotonically decreasing
we have that the limit of the sequence has the lower bound
\beq f_\infty>1-\sum_{n=1}^\infty {\cal A}^{n,n}\label{finflwrbnd}\eeq
which is positive at
 large enough $\mu$. On the other hand, as 
$\mu$ decreases the sequence will go negative and then start to oscillate. It
is clear that at  small enough $\mu$ this must happen because it is
straightforward to choose a $\mu$ for which $f_2$ is already negative. It
follows
by continuity  that 
the critical point where the sequence $\{f_\ell\}$ monotonically decreases
to $f_\infty=0$ must exist, and therefore the system is a branched polymer.

\begin{figure}[htb]
\epsfxsize=4.2in
\bigskip
\centerline{\epsffile{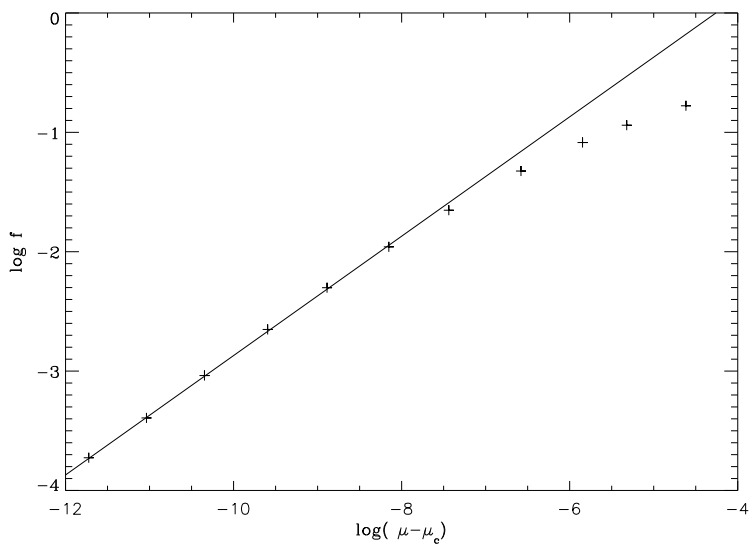}}
\caption{\baselineskip=12pt {\it Logarithmic plot of $f_\infty$ against $\mu$ 
for the case $w(n)=1$, $\e^{-\kappa}=10^{-4}$. The crosses represent computed
values with $\mu_c=0.040112$ and the straight line has
slope $\frac12$}.}
\bigskip
\label{gammaplot}\end{figure}

To determine the value of $\gstr$ we consider two cases. If $w(n)$ is
a strictly increasing function of $n$, then by working at very large $\kappa$
we can ensure that the first term in \eqn{BPcondn} dominates so that at the
critical point
\beq 1\simeq\Bigl(g_1(\bar\mu_1)-g_1^{\prime}(\bar\mu_1)\Bigr)
{}~\e^{-\kappa}\eeq
and higher terms are suppressed by factors of $\e^{-\kappa(w(n)-w(1))}$.
As long as $\{f_\ell\}$ is a decreasing sequence, the sum of terms with
$n\ge 2$ in \eqn{BPcondn} converges to something negligible compared to the
first term.
Now we can make the standard argument that the denominator of \eqn{dZ1eqn} will
vanish linearly with $\widehat{\cal Z}_1(\mu,\kappa)$ which we assume to behave
as
\beq
\widehat{\cal Z}_1(\mu,\kappa)=\widehat{\cal
Z}_1\Bigl(\mu_c(\kappa),\kappa\Bigr)
-B\Bigl(\mu_c(\kappa),\kappa\Bigr)\,\Bigl(\mu-\mu_c(\kappa)\Bigr)^{1-\gstr}
\eeq
where the function $B$ is regular at the critical point, and therefore
$\gstr=\frac12$.
In the case that $w(n)$ is constant this argument fails. To show that the
model
has generic branched polymer behaviour we would need to show that 
\beq \left.\frac{\partial^2{\cal F}}{\partial\widehat{\cal
Z}_1^2}\right)_{\mu,\kappa}\ne 0~~~~~~\hbox{when}~~f_\infty=0\eeq
but we have not succeeded in finding a proof of this. However, it is 
straightforward to solve the recursion equations \eqn{flrecrel} numerically and
fig. \ref{gammaplot}
shows the behaviour of $f_\infty$ for $\e^{-\kappa}=10^{-4}$. It is clear 
that $\gstr=\frac12$ in this case as well.

We conclude that the system defines a generic branched polymer phase of two
dimensional
quantum gravity with string susceptibility exponent $\frac12$ as soon as 
the coupling  $\kappa$ is tuned away from infinity. This model only allows
a rather simple subset of all possible branchings and we expect that
 the ordinary pure gravity phase (with
$\gstr=-\frac12$) is recovered by including all of the more complicated
branchings
in which the path on the base graph is allowed to be arbitrary.
The basic point is that the model above gives a physical understanding 
for the absence of a flat phase when the curvature coupling constant $\kappa$
is large.

\vskip 1cm

This work was supported in part by PPARC grant \# GR/L56565 and the Danish
Natural Science Research Council.

\end{document}